\documentclass{optica-article}

\journal{opticajournal} 

\articletype{Research Article}
\usepackage{float}
\usepackage{hyperref}
\usepackage{lineno}

\begin{document}

\title{Inverse Design of Photonic Crystal Waveguides Using Neural Networks and Dispersion Optimization}

\author{Lyucheng Feng\authormark{*}}

\address{Department of Electronic \& Electrical Engineering, University College London, Torrington Place, London WC1E 7JE, UK}

\email{\authormark{*}zceelfe@ucl.ac.uk} 


\begin{abstract*} 
Photonic crystal waveguides (PCWs) play a critical role in precisely controlling light propagation, enabling high-performance functions in applications such as optical communication and integrated photonics. The design of PCWs traditionally relies on complex numerical methods, including finite-difference time-domain (FDTD) and plane-wave expansion (PWE) methods, which are often inefficient when dealing with high-dimensional parameter spaces, particularly for subwavelength structures. To overcome these challenges, a convolutional neural network (CNN)-based inverse design method is introduced to optimize the structural parameters of PCWs. By simulating band structures under varying line defect widths and air hole radii using MIT Photonic Bands (MPB), a large dataset was generated, mapping structural parameters to corresponding band characteristics. Backpropagation neural networks (BPNN) and CNN models were trained on this dataset to predict key PCW structural parameters. The CNN model demonstrated superior performance in predicting complex geometries, maintaining high accuracy even when extrapolating beyond the training dataset, with precision up to four decimal places. In contrast, the BPNN model exhibited faster training times and higher computational efficiency on smaller datasets, though it performed less effectively on larger datasets. Cross-validation using MPB confirmed the generalization capability and reliability of both models. This study highlights the potential of deep learning techniques in photonic device design, particularly for advancing the development of high-efficiency, low-loss components in integrated photonics and optical communication systems.

\end{abstract*}

\section{Introduction}
Photonic crystals (PCs) are engineered optical materials characterized by periodic variations in dielectric constant. Their defining feature is the presence of photonic bandgaps (PBGs), within which no extended states are permitted, and the incident light is reflected. By leveraging this property, a line defect can be introduced into a two-dimensional photonic crystal structure, the fundamental concept involves carving a waveguide by modifying a linear sequence of unit cells within the otherwise perfect photonic crystal, thereby confining and directing light with frequencies within the PBG along the defect\cite{Joannopoulos:08:Book}. Consequently, photonic crystal waveguides (PCWs) not only exhibit superior light-guiding properties but also deliver outstanding performance in various cutting-edge optical applications. 

Line-defect photonic crystal waveguides offer remarkable advantages in slow light applications. By precisely optimizing their structure, it is possible to effectively reduce the group velocity and manage group velocity dispersion (GVD), making these waveguides particularly valuable for optical signal buffering and similar applications\cite{2,3,4}. Moreover, in nonlinear optics, optimizing the geometry and composition can significantly enhance nonlinear interactions. For instance, in high core-cladding index contrast waveguides, there is an optimal waveguide size that maximize the non linear coefficient, enhancing light confinement and reducing power needed for nonlinear interactions, such as supercontinuum generation, supporting the development of ultra-low-power optical devices\cite{5}. Furthermore, highly nonlinear slot waveguides, which achieve flat and low-dispersion characteristics, significantly improve the potential for efficient optical signal processing, particularly by enhancing the nonlinear coefficient, which is critical for various on-chip applications\cite{6}. In recent years, the field of topological photonics has further expanded the application scope of PCWs. Utilizing topological protection mechanisms, such as the spin-orbit coupling that enables unidirectional edge state propagation, these waveguides can achieve backscattering-free light transmission, thereby significantly reducing optical losses and enhancing device stability and reliability\cite{7}. In the context of high-speed optical communications, structural optimization of waveguides plays a critical role in dispersion compensation. For instance, the design of dual-waveguide structures has been employed to compensate for GVD, thereby improving the quality and integrity of signal transmission\cite{8}.

However, designing PCWs requires precise control over their dispersion characteristics to achieve specific optical performance. This often involves fine-tuning the geometric structure of the waveguide, such as the width of the line defect waveguide and the radius of the air holes\cite{9}. Traditional design methods primarily rely on numerical simulations, such as the FDTD method or FEM. While these methods can provide accurate results, they require high-resolution meshing of the entire structure and involve solving complex optical equations iteratively to obtain dispersion relations and other optical characteristics. Particularly when multi-parameter optimization is simultaneously involved, this forward design process demands significant computational resources and time, making it challenging to determine optimal design parameters within a short timeframe\cite{10,11}.

Given these challenges, inverse design methods have emerged in recent years within photonics, demonstrating considerable advantages. Inverse design has rapidly gained popularity in nanophotonic, driven by the demand for advanced photonic devices. This approach has led to the development of ultra-fast, data-driven solvers for complex inverse design challenges\cite{12,13}.Particularly effective for optimizing intricate photonic structures, inverse design has been used to enhance the dispersion and quality factor of photonic crystal waveguides and ultrasmall cavities\cite{14}, thereby broadening the scope of applications in sophisticated optical systems. Additionally, it has enabled the development of non-reciprocal pulse routers for chip-scale LiDAR systems, improving their performance by reducing insertion loss and enhancing power handling capabilities\cite{15}. Furthermore, inverse design has facilitated the creation of compact, broadband on-chip wavelength demultiplexers, which boost efficiency and minimize crosstalk\cite{16}. Collectively, these advancements underscore the potential of inverse design in the miniaturization and optimization of integrated photonic devices.

With the impressive performance of inverse design above, it allows for the direct prediction of the necessary structural parameters based on desired optical performance targets, such as specific dispersion relations, effectively bypassing the need for extensive forward calculations. This is particularly valuable in the design of photonic crystals waveguides and other complex photonic devices, where the optimization often involves multiple parameters (such as line defect waveguide width and air hole radius) that are complexly and nonlinearly interconnected. Traditional optimization methods may encounter computational challenges in handling these issues. Inverse design, leveraging deep learning techniques, such as Convolutional Neural Network, effectively navigates these complex multi-parameter spaces and directly produce parameter combinations that meet the requirements through model training. This method not only enhances design efficiency but also reduces computational time, offering a powerful alternative to traditional simulation methods and enabling the development of more advanced optical systems.

Machine learning techniques, particularly artificial neural networks (ANNs), have been successfully employed to correct inaccuracies in fiber nonlinearity models, such as the Gaussian-noise model, improving the precision of nonlinear interference (NLI) estimation. Combined with dynamic monitoring schemes, these ML methods provide more accurate NLI variance assessments\cite{17}. Moreover, In photonic device design, deep neural networks (DNNs) predict and optimize key components like silicon photonic grating couplers, significantly boosting computational efficiency and device performance\cite{18}. For nanophotonic devices, deep learning models incorporating physical constraints, like the sparse parameter method (SPM), have shown strong generalization capabilities, enabling accurate forward predictions and multi-functional inverse design across different wavelengths and polarizations, optimizing device performance with minimal retraining\cite{19}. The effectiveness of this approach has also been demonstrated in a wide range of applications, including  topological photonic structures, as well as near-field and on-chip optics\cite{20,21,22,23,24,25}.

The primary objective of this research is to address an engineering problem through the development of an inverse design method based on CNN and BPNN for optimizing the structural parameters of PCWs with line defects. Specifically, we will utilize the MPB software to design and simulate the band structures under varying conditions of line defect width and air hole radius. By processing these band structure data, we will generate a large dataset that maps structural parameters to their corresponding band characteristics, which will serve as the training foundation for both the CNN and BPNN models. Ultimately, the trained CNN and BPNN models will be employed to achieve inverse design of PCW structural parameters given specific dispersion relationships. This approach not only significantly reduces the computational time and resources required for designing complex photonic structures but also enhances design accuracy and robustness. 

The remainder of the paper is organized as follows: In Section 2, the methodology for the neural network-based inverse design is introduced, covering the forward and inverse design formulas, MPB modeling, and data collection processes, along with model validation through FDTD simulations. This section also provides an overview of the CNN and BPNN models, detailing the training process and hyperparameter optimization. In Section 3, the simulation results are presented, focusing on the comparison of CNN and BPNN performance, as well as the accuracy of inverse design verified through MPB cross-validation. Section 4 concludes the study with a summary of key findings.

\section{Data Collection and Deep Learning Model}
\subsection{Methodology Overview}
This study proposes and develops a neural network-based inverse design method aimed at optimizing the structural parameters of PCWs. The method begins by constructing a model using MPB software, employing a triangular lattice to control the distribution of PBGs. To reduce computational errors due to periodic boundary conditions, a supercell design is applied. The key parameters, air hole radius (r) and waveguide width (w), are adjusted, and the corresponding dispersion relations are computed and analyzed to capture the optical characteristics of the structure.

To validate the accuracy of the constructed model, spectral simulations were performed using the FDTD method. PBGs characteristics and loss variations were monitored to ensure that the designed structures met the expected performance requirements. Specifically, by adjusting the center frequency, the consistency of the dispersion relations with the MPB simulation results was examined, further confirming the model's reliability. Based on the validated model, a large dataset was generated by simulating various combinations of r and w parameters. These simulations yield corresponding dispersion relations, forming the dataset for subsequent neural network training.

During the neural network training phase, the TE mode frequency data extracted from MPB simulations were preprocessed to make them suitable for neural network input. This included normalization and standardization to meet the training requirements. Two models, a BPNN and a CNN were then constructed and trained using the processed structural parameters (e.g. air hole radius r and waveguide width w) and their associated dispersion relations.  After training, the models and their results were saved. A comparative performance analysis was conducted to assess the prediction accuracy of both models, with hyperparameters such as learning rate and the number of neurons in the hidden layer being fine-tuned to optimize performance.

Once the training was complete, cross-validation was carried out. Two groups of artificial designed photonic crystal structures with different waveguide defect modes were designed, and the trained models were used for prediction. The predicted structural parameters were then re-simulated using MPB to generate the corresponding dispersion relations, which were compared with the neural network predictions. This process verified the accuracy of the model’s predictions and its reliability in practical applications. 
\begin{figure}[H]
    \centering
    \includegraphics[width=1\linewidth]{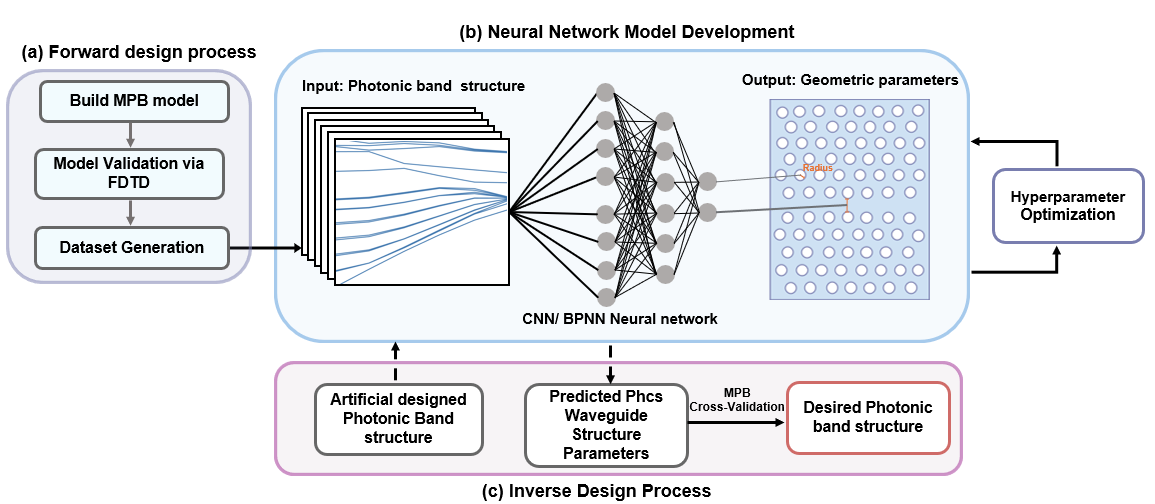}
    \caption{Overview of the neural network-based optimization process for the inverse design of photonic crystal waveguides. (a) Forward design process.     
     (b) Training process of the neural network model, which processes input photonic band structures to predict air hole radius and waveguide width as outputs. (c) The procedures in the inverse design and cross-validation of waveguide predicted parameters.}
    \label{fig:enter-label}
\end{figure}

\subsection{Forward Process and Inverse Design Formulas}
The core of the forward design process is to calculate the dispersion relations based on known geometric parameters and to generate a corresponding dataset. Following this, the inverse design derives the optimal geometric parameters by back-calculating from a given target frequency.

In the forward design process, it is first necessary to define the range of design parameters, primarily including the air hole radius $r$ and waveguide width $w$. In this study, the air hole radius ranges from 0.15 to $0.49 \mu \mathrm{m}$, while the waveguide width varies from 1.0 to $1.98 \mu \mathrm{m}$. For each combination of parameters $\left(r_k, w_k\right)$, the forward model is executed to calculate the corresponding dispersion relations. Equation (1) presents this relationship:
\begin{equation}
    \omega_k=f\left(r_k, w_k\right)
\end{equation} 

Where $\omega_k$ represents the dispersion characteristics at different frequencies.
Each resulting set of parameters $\left(r_k, w_k, \omega_k(k)\right)$ is stored as a dataset $D_{f w}^k$, shown in Equation (2):
\begin{equation}
    D_{f w}^k=\left\{\left(r_k, w_k, \omega_k(k)\right)\right\}
\end{equation} 

Additionally, a band structure analysis is performed on this dataset. Through forward design, the influence of geometric parameters on the dispersion characteristics of photonic crystals is explored, providing the foundational data for subsequent inverse design.

In the inverse design process, the dataset, which is generated based on the results of the forward design model, serves as the foundation. This dataset is used to train the inverse neural network model $N N_{\text {inv }}$, which predicts the corresponding geometric parameters $\left(r_{\text {targ }}, w_{\text {targ }}\right)$ given a target dispersion relation $\omega_{\text {targ }}$. The predictive relationship is shown in Equation (3):
\begin{equation}
    r_{\text {targ }}, w_{\text {targ }}=N N_{\text {inv }}\left(\omega_{\text {targ }}\right)
\end{equation} 

The predicted geometric parameters are subsequently fed into the forward model to calculate the corresponding dispersion relation, as indicated in Equation (4):
\begin{equation}
    \omega_{\text {pred }}=N N_{\text {fw }}\left(r_{\text {targ }}, w_{\text {targ }}\right)
\end{equation} 

To evaluate the accuracy of the model, the error $e$ between the target frequency $\omega_{\text {targ }}$ and the predicted frequency $\omega_{\text {pred }}$ is determined, as described by Equation (5):
\begin{equation}
    e=\operatorname{Err}\left(\omega_{\text {targ }}, \omega_{\text {pred }}\right)
\end{equation}

\subsection{MPB Modeling and Data Collection}
MPB is an open-source software tool designed for calculating the band structure of photonic crystals. It is based on the PWE, allowing efficient computation of dispersion relations and photonic bandgaps for various photonic crystal structures. In this study, MPB was employed to simulate the optical properties of photonic crystal waveguides.

To achieve effective bandgap formation, we selected a triangular lattice photonic crystal structure characterized by high symmetry and uniformity\cite{Joannopoulos:08:Book}. This structure demonstrates stable photonic bandgap characteristics, making it particularly suitable for photonic waveguide design. To further investigate the influence of waveguide width on bandgap properties, a supercell structure was introduced. The supercell size was set as $13+2 x$, where $x$ represents the variation in waveguide with on one side. By dynamically adjusting the supercell size, we ensured that the simulation region was sufficiently large to minimize boundary effects.

The key parameters in the model were the air hole radius $r$ and waveguide width $w$, which varied from 0.15 to $0.49 \mu \mathrm{m}$ and from 1.0 to $1.98 \mu \mathrm{m}$, respectively. Through fine-tuning with an interval of 0.01 , a total of 1803 datasets were generated. These datasets were used to analyze the distribution of photonic bandgaps and guided modes. As the waveguide width changed, the positions of the air holes were also precisely adjusted. To ensure structural consistency and uniformity, the positions of the air holes were manually defined. With varying waveguide widths, the air hole positions were dynamically adjusted through calculated displacements $d x$, ensuring that the layout within the lattice remained precise and maintained symmetry in the simulated structure, as illustrated in Fig.\ref{fig:2}:

\begin{figure}[H]
    \centering
\begin{minipage}{.49\linewidth}
    \centering
    \includegraphics[width=\linewidth]{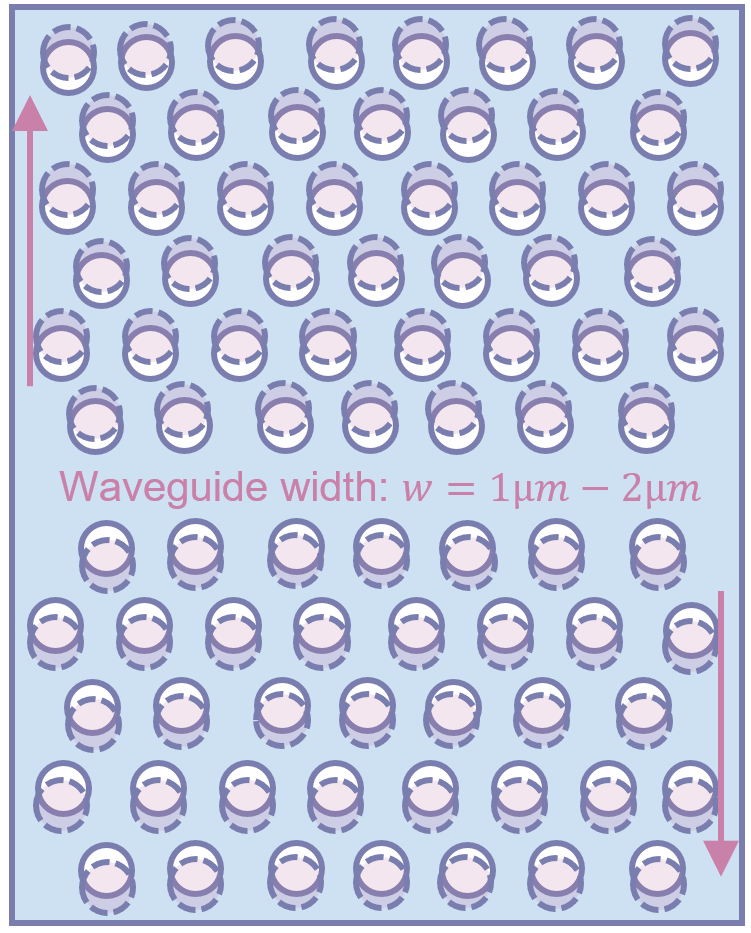}\\
    (a)
\end{minipage}

\begin{minipage}{.49\linewidth}
    \centering
    \includegraphics[width=\linewidth]{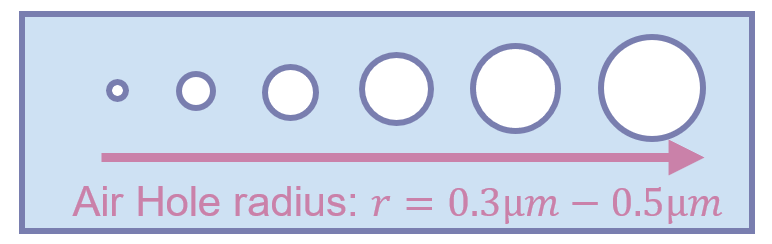}\\
    (b)
\end{minipage}

    \caption{Model of a triangular lattice photonic crystal waveguide constructed in MIT Photonic Bands (MPB) with a default material permittivity of $\epsilon=12$, illustrating variable parameters. (a) Displacement of air holes corresponding to changes in waveguide width, ranging from 1.0 $\mu$m to 1.98 $\mu$m. 
    (b) Gradual increase in air hole radius from 0.15 $\mu$m to 0.49 $\mu$m, in increments of 0.01 $\mu$m.}
    \label{fig:2}
    
\end{figure}

Periodic boundary conditions (PBC) were applied in the MPB simulations to mimic an infinite photonic crystal structure. These boundary conditions were set as periodic in both the $x$ and $y$ directions, allowing for the accurate simulation of the infinitely repeating nature of the photonic crystal lattice. This setup enabled the precise capture of guided mode patterns and photonic bandgap properties within a finite supercell, while minimizing the influence of boundary effects.

For analyzing the dispersion relations, the projected band diagram was employed. This diagram illustrates the dispersion relation along a specific direction with one-dimensional periodicity in the lattice. Since the lattice is periodic along this direction, the wave vector $k$ remains conserved. In our study, A path was selected from the center of the Brillouin zone ( $\Gamma$ point) to the boundary ( $K^{\prime}$ point), and four $k$-points were interpolated. This approach facilitated the efficient capture of the photonic crystal's dispersion relations while maintaining computational efficiency.

Additionally, during the band structure calculations performed using MPB simulations, 18 bands were set to capture both guided mode behavior and higher-order band characteristics. Upon completing the simulations, the dielectric constant distribution was extracted, and visualizations were generated to illustrate the photonic bandgap behavior under different structural parameters. The analysis focused on the position and width of the bandgaps, providing essential data for optimizing the design of photonic crystal waveguides.

\subsubsection{Model Validation with FDTD}
To validate the accuracy of the MPB simulation results, the open-source software Meep, based on the FDTD method, was utilized. In this study, the fundamental dispersion relations and bandgap positions of the photonic crystal waveguide were first determined using MPB. To ensure the reliability of the MPB results, an identical model was constructed in Meep, and the flux intensity in different regions was compared to validate the findings.   
\begin{figure}[H]
    \centering
\begin{minipage}{.49\linewidth}
    \centering
    \includegraphics[width=\linewidth]{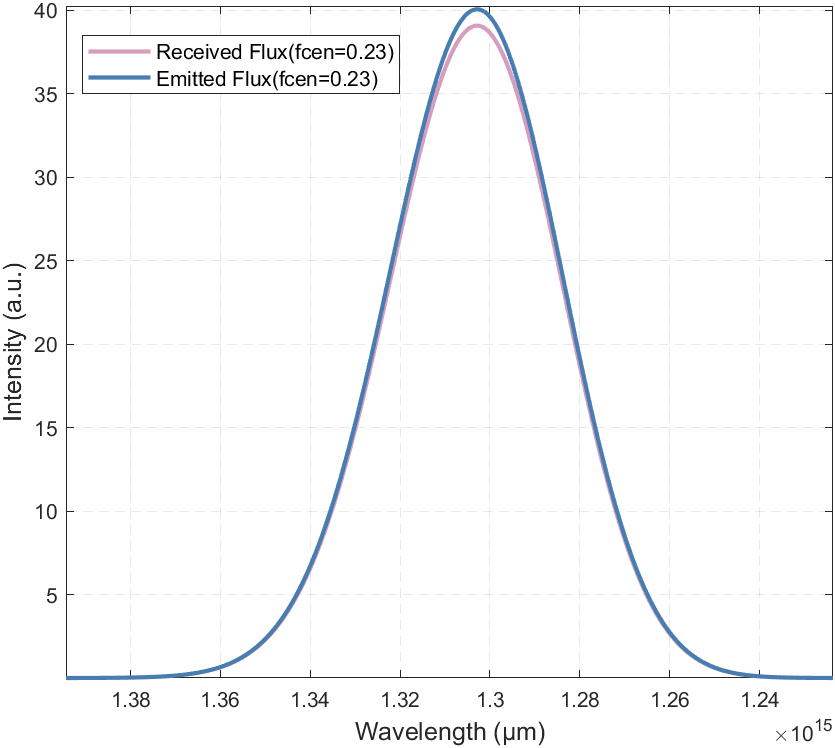}\\
    \hypertarget{fig:3a}{(a)}
\end{minipage}

\begin{minipage}{.49\linewidth}
    \centering
    \includegraphics[width=\linewidth]{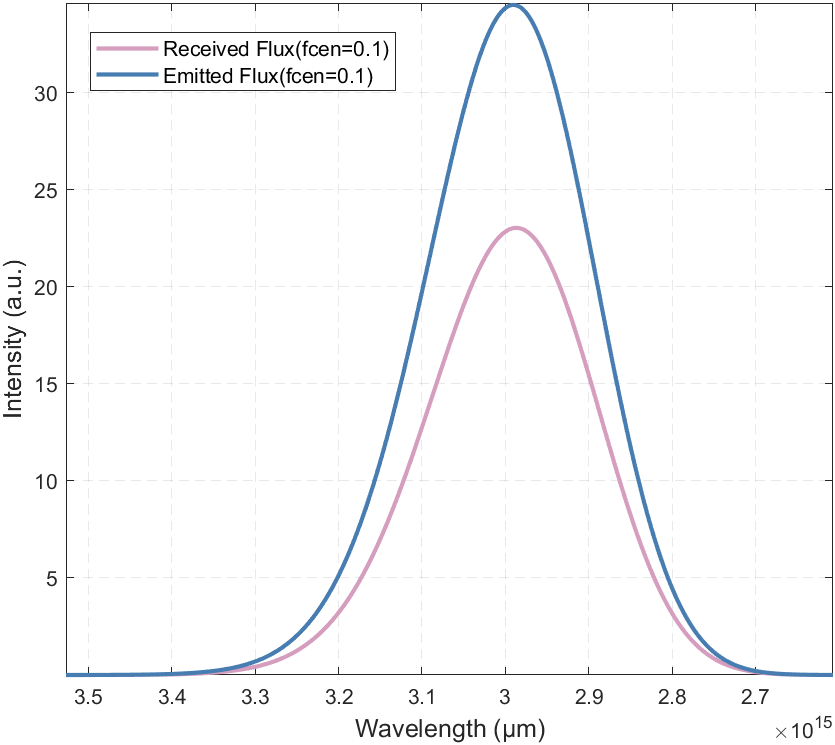}\\
      \hypertarget{fig:3b}{(b)}
\end{minipage}
 \caption{Validation of MPB model accuracy via FDTD-based Meep analysis across varying \( f_{\text{center}} \) conditions for photonic crystal waveguides. (a) Flux intensity profiles at a center frequency \( f_{\text{center}} = 0.23 \) within the bandgap. (b) Increased energy loss observed at a center frequency \( f_{\text{center}} = 0.1 \), indicating conditions outside the bandgap. In both (a) and (b), the blue line represents emitted flux and the red line represents received flux.}

\end{figure}

During the Meep simulations, all parameters were kept consistent with those used in the MPB simulations. By adjusting the center frequency of the emitted wave $f_{\text {center }}$, the energy loss at different frequencies was analyzed. As shown in Fig.3\hyperlink{fig:3a}{(a)}, when the emission frequency was within the bandgap range, the energy loss was minimal (for a center frequency of $f_{\text {center }}=0.23$, the energy loss was only $2.43 \%$ ). In contrast, Fig.3\hyperlink{fig:3b}{(b)} shows when the frequency was outside the bandgap range, energy loss increased significantly (for a center frequency of $f_{\text {center }}=0.1$, the energy loss reached $33.27 \%$ ).

\begin{figure}[H]
    \centering
\begin{minipage}{.49\linewidth}
    \centering
    \includegraphics[width=\linewidth]{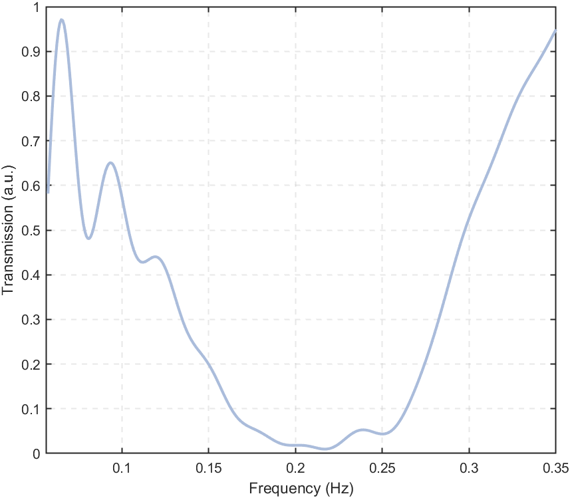}\\
    (a)
\end{minipage}
\begin{minipage}{.49\linewidth}
    \centering
    \includegraphics[width=\linewidth,height=5.6cm]{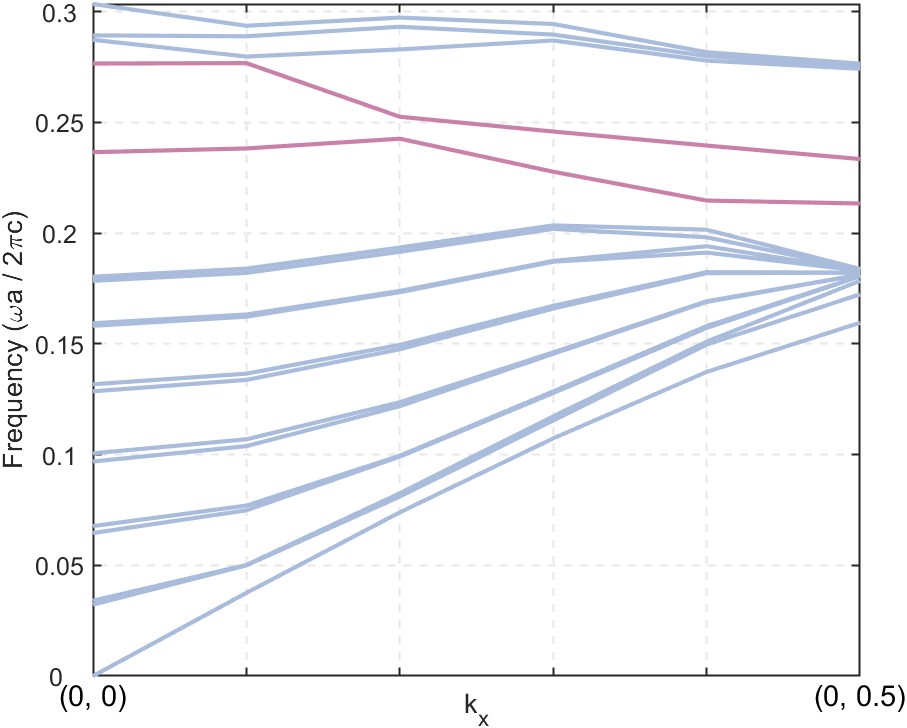}\\
    (b)

\end{minipage}
\caption{Further validation of MPB model accuracy through FDTD-based Meep simulations. (a) Transmission spectrum showing the suppression of light propagation within the bandgap frequencies due to air hole structures. (b) Band diagram generated by MPB, illustrating the bandgap across different wave vectors \(k_x\), with the red line representing the guided band within the bandgap.}
    \label{fig:4}
\end{figure}

Meep was also used to simulate the flux comparison of the photonic crystal waveguide with and without air hole structures, generating the corresponding transmission spectrum. As illustrated in Fig.\ref{fig:4}, the photonic crystal waveguide with air holes effectively suppressed light propagation within the bandgap range, which was highly consistent with the bandgap characteristics of the photonic crystal. These results were in complete agreement with the bandgap phenomena observed in the MPB-generated band diagrams, further validating the accuracy and reliability of the model.

\subsection{Neural Network Model Development: CNN and BPNN}
In this study, both BPNN and CNN were constructed and employed to predict the structural parameters of photonic crystal waveguides, specifically the air hole radius r and waveguide width w. These models were utilized to learn and predict the mapping relationship between photonic bandgap characteristics and the structural parameters.

The BPNN is a classical multilayer perceptron (MLP) model capable of mapping input parameters to output results. The architecture consists of an input layer with 108 nodes, a hidden layer with 9 neurons, and an output layer with 2 nodes, corresponding to the two structural parameters depicted in Fig.5\hyperlink{fig:5a}{(a)}. The input layer represents the 108 features of each sample, and the hidden layer employs a tansig activation function (hyperbolic tangent activation function), while the output layer uses a purelin activation function (linear activation function),As illustrated in Fig.5\hyperlink{fig:5b}{(b)}. The model is trained using the mean squared error (MSE) as the loss function to evaluate prediction accuracy.
\begin{figure}[H]
    \centering
\begin{minipage}{.49\linewidth}
    \centering
    \includegraphics[width=0.8\linewidth]{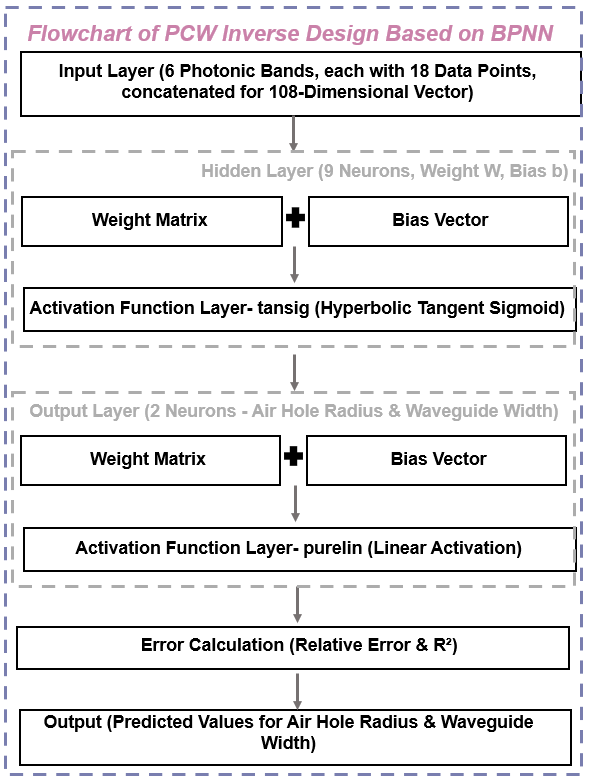}\\
    \hypertarget{fig:5a}{(a)}
\end{minipage}
\begin{minipage}{.49\linewidth}
    \centering
    \includegraphics[width=\linewidth]{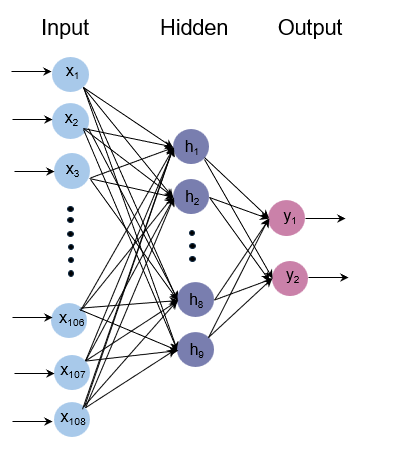}\\
    \hypertarget{fig:5b}{(b)}
\end{minipage}

    \caption{Structure of the BPNN model for the inverse design of PCW. (a) Flowchart of the BPNN inverse design process, outlining the main layers and connections. (b) Schematic representation of the neural network architecture, showing the input, hidden, and output layers.}

    \label{fig:enter-label}
\end{figure}

\begin{figure}[H]
    \centering
    \includegraphics[width=0.4\linewidth]{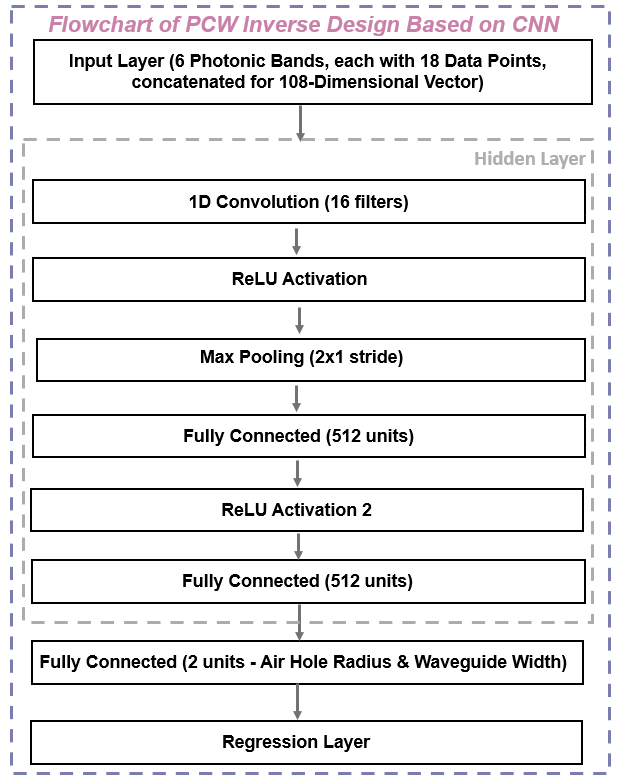}\\
    (a) \\
    \includegraphics[width=\linewidth]{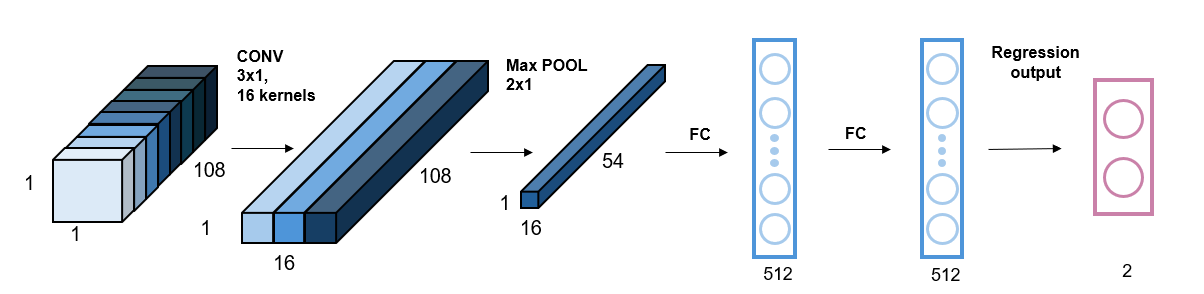}\\
    (b)
    \caption{Structure of the CNN model used for the inverse design of PCW. (a) Flowchart of the CNN-based inverse design process, outlining the main layers and activation functions. (b) Schematic representation of the CNN architecture, illustrating the flow from the input layer through the convolutional and fully connected layers to the final regression output.}
    \label{fig:6}
\end{figure}

The CNN is a classical deep learning model, which excels at handling spatial data with local dependencies. In this study, the CNN architecture includes an input layer, a convolutional layer, a pooling layer, two fully connected layers, and an output layer. The convolutional layer utilizes a 3×1 kernel with a stride of 1×1 and applies 16 filters to extract local features. The feature maps are then passed through a ReLU activation function to introduce non-linearity.  Subsequently, a max-pooling layer with a 2×1 pooling window is used to down-sample the feature maps, reducing their dimensionality while retaining critical information. The resulting features are subsequently fed into two fully connected layers, each containing 512 neurons, to further process the extracted features, as shown in Fig.\ref{fig:6}. Finally, the output layer consists of two neurons that predict the air hole radius and waveguide width. Like the BPNN, the CNN model is trained using MSE as the loss function to evaluate prediction errors.

\subsubsection{Training Process and Hyperparameter Optimization}
During the training process of both neural network models, the dataset was randomly divided into a training set and a test set. A total of 1796 samples were randomly selected from the 1803 available samples for training, while the remaining 7 samples were used for testing. This division ensured diversity within the training set while maintaining the independence of the test set, thereby enhancing the model's generalization capability.

The BPNN model was trained using the standard backpropagation algorithm, with weight updates performed by the Levenberg-Marquardt algorithm (trainlm). After normalization, the input data were fed into the network, and the output was calculated via forward propagation. The MSE loss function was employed to compute the error between the predicted and actual values, and the network weights were adjusted through backpropagation. The training process was configured with a maximum of 1000 iterations and a target error of 0.001 (MSE). To prevent overfitting, a regularization factor of 0.1 was applied, and the learning rate was set to 0.001. These hyperparameters, including the number of hidden layer neurons, were fine-tuned through experiments. A total of 9 neurons in the hidden layer provided the optimal balance between feature extraction and overfitting, while the learning rate of 0.001 ensured stable training. The regularization factor of 0.1 further enhanced generalization, reducing overfitting.

The CNN model followed a similar training process, using forward propagation to compute the output and MSE as the loss function. The Adam optimizer was employed to adjust the weights during backpropagation to minimize the loss. The training process was configured with a maximum of 200 iterations, an initial learning rate of 0.001, and the mini-batch method (batch size of 12) to accelerate training and reduce the risk of overfitting. The tuning of CNN hyperparameters included the size of the convolutional kernel, pooling layer stride, the number of neurons in fully connected layers, and the learning rate. Experimental results demonstrated that a 3×1 convolutional kernel effectively extracted local geometric features, while 16 filters ensured sufficient feature extraction. The pooling layer stride was set to 2×2, compressing the feature map while retaining critical information. Two fully connected layers, each with 512 neurons, balanced model complexity and learning capacity. A learning rate of 0.001 ensured stable convergence during training. Additionally, the introduction of regularization and the mini-batch method further reduced overfitting, enhancing the model's generalization.

\section{Simulation Results and Discussion}
\subsection{Analysis of Training Outcomes}
During training, the CNN model exhibited strong convergence. The initial loss decreased rapidly within the first iteration, with the final MSE reaching 0.018507 and the root mean squared error (RMSE) reaching 0.0058205. Throughout the training process, the validation loss remained consistent with the training loss, indicating no overfitting and demonstrating the model's robust regression prediction capability. 

\begin{figure}[H]
    \centering
\begin{minipage}{.49\linewidth}
    \centering
    \includegraphics[width=\linewidth]{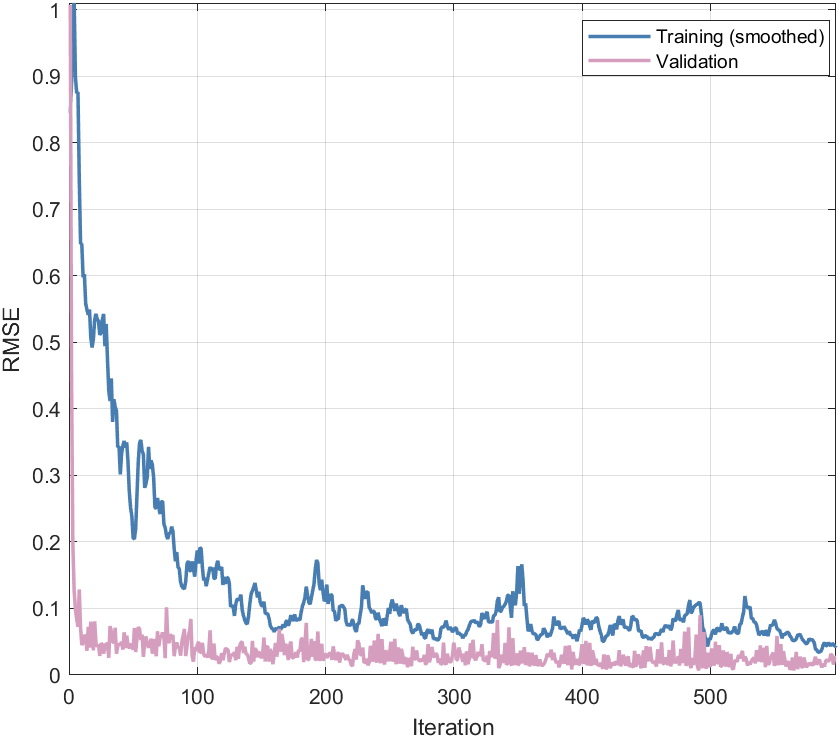}\\
    (a)
\end{minipage}
\begin{minipage}{.49\linewidth}
    \centering
    \includegraphics[width=\linewidth]{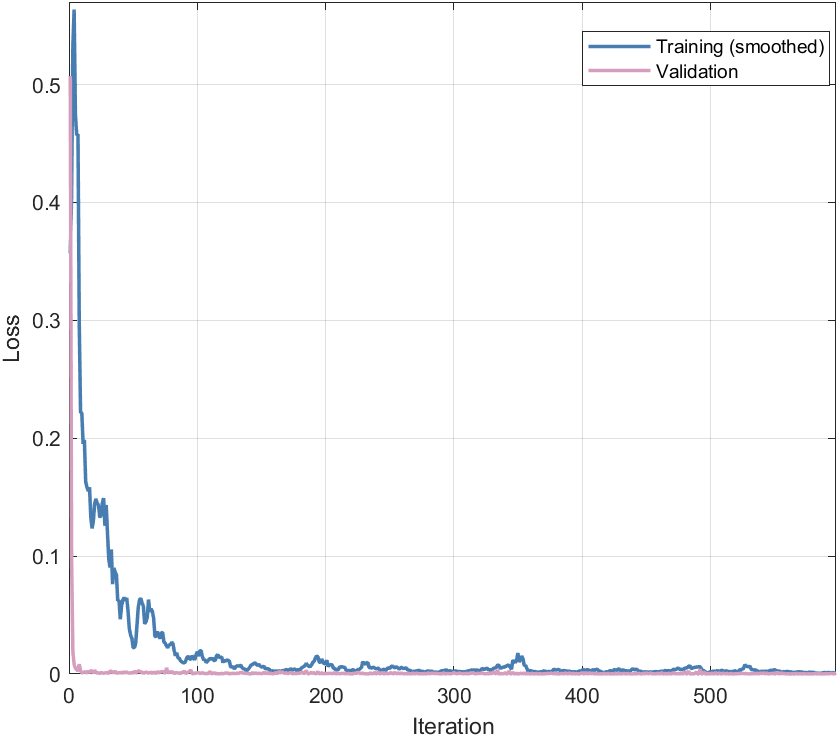}\\
    (b)
\end{minipage}
   \caption{Loss and RMSE curves of the CNN model during training and validation. (a) RMSE values for training (blue line) and validation (red line) across 600 iterations. (b) Loss curves for training (blue line) and validation (red line) over the same 600 iterations.}

    \label{fig:7}
\end{figure}

The loss and RMSE curves clearly illustrate the model’s learning trend and convergence process, as depicted in Fig.\ref{fig:7}. Particularly in the first epoch (with 150 iterations per epoch), the RMSE dropped from 1 to 0.1, and the loss value decreased from nearly 0.6 to approximately 0.01, highlighting the model's rapid convergence ability.

The BPNN model also showed stable training performance. During the first 8 epochs, the loss function rapidly decreased from 1.0 to 0.010379, reaching optimal performance in the 7th epoch, as illustrated in Fig.8\hyperlink{fig:8a}{(a)}. Similarly, no overfitting was observed in the BPNN model. Compared to CNN, the BPNN model converged faster when handling smaller datasets. The gradient convergence performance indicated that the gradient reached 0.378 in the 5th epoch, suggesting that the model was approaching convergence, with the training process remaining stable, as evidenced by Fig.8\hyperlink{fig:8b}{(b)}. Furthermore, the regression analysis for the training, validation, and test sets, presented in Fig.8\hyperlink{fig:8c}{(c)}, confirmed a strong linear correlation between the predicted and actual values, with R-values close to 1, underscoring the model’s predictive accuracy.
\begin{figure}[H]
    \centering
\begin{minipage}{.49\linewidth}
    \centering
    \includegraphics[width=\linewidth]{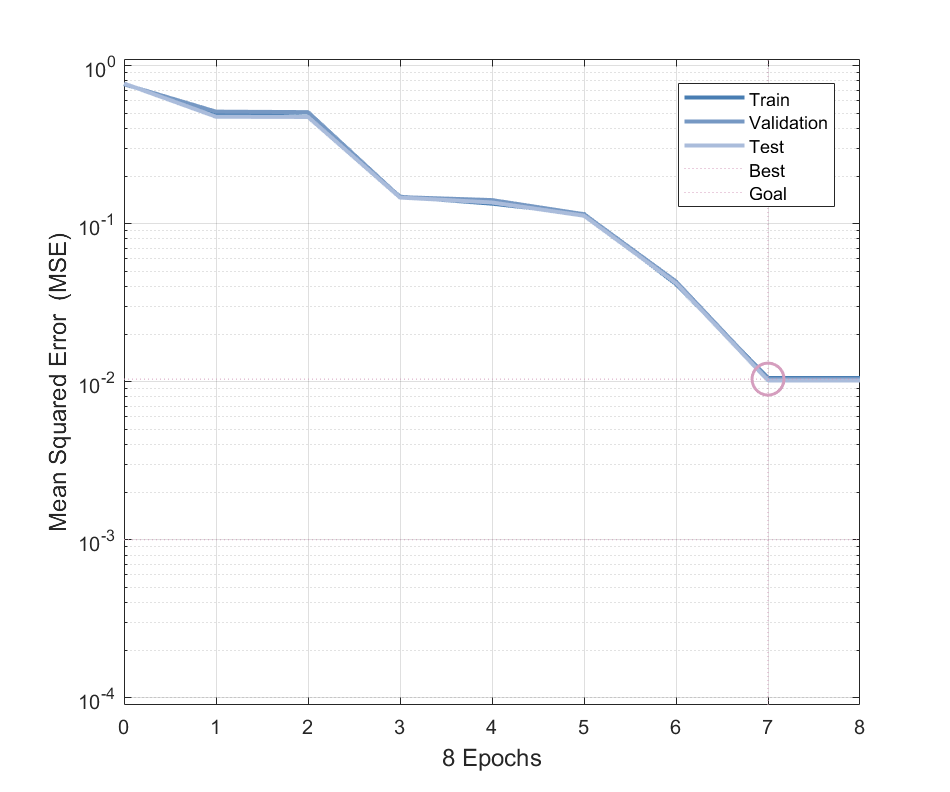}\\
    \hypertarget{fig:8a}{(a)}
\end{minipage}
\begin{minipage}{.49\linewidth}
    \centering
    \includegraphics[width=\linewidth]{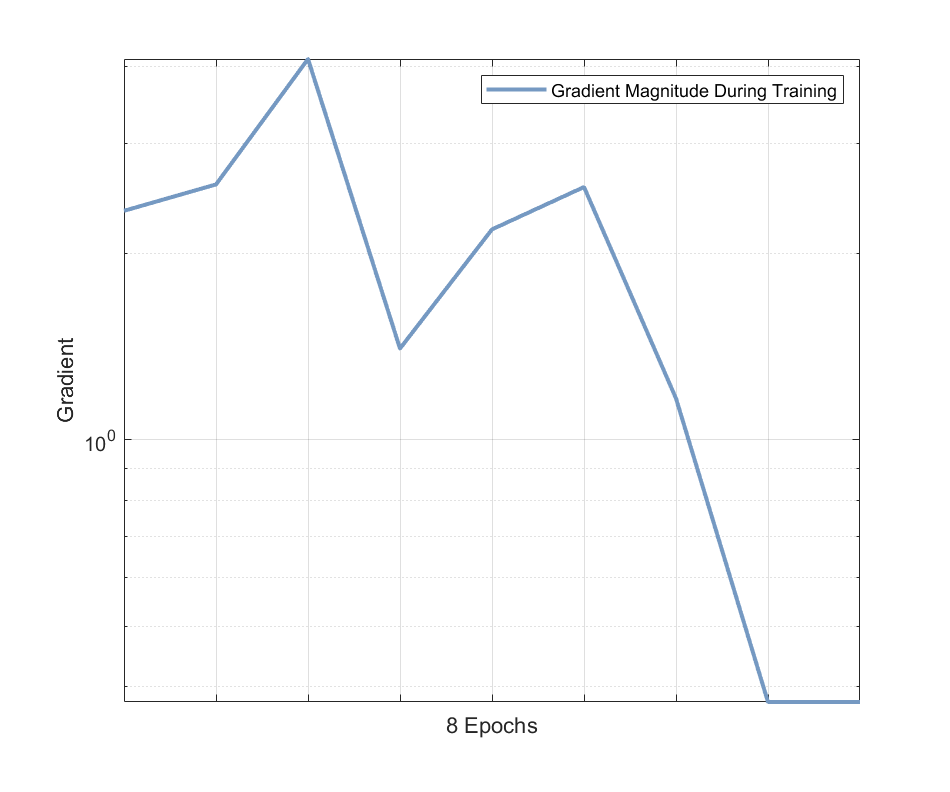}\\
        \hypertarget{fig:8b}{(b)}
\end{minipage}

\begin{minipage}{.49\linewidth}
    \centering
    \includegraphics[width=\linewidth]{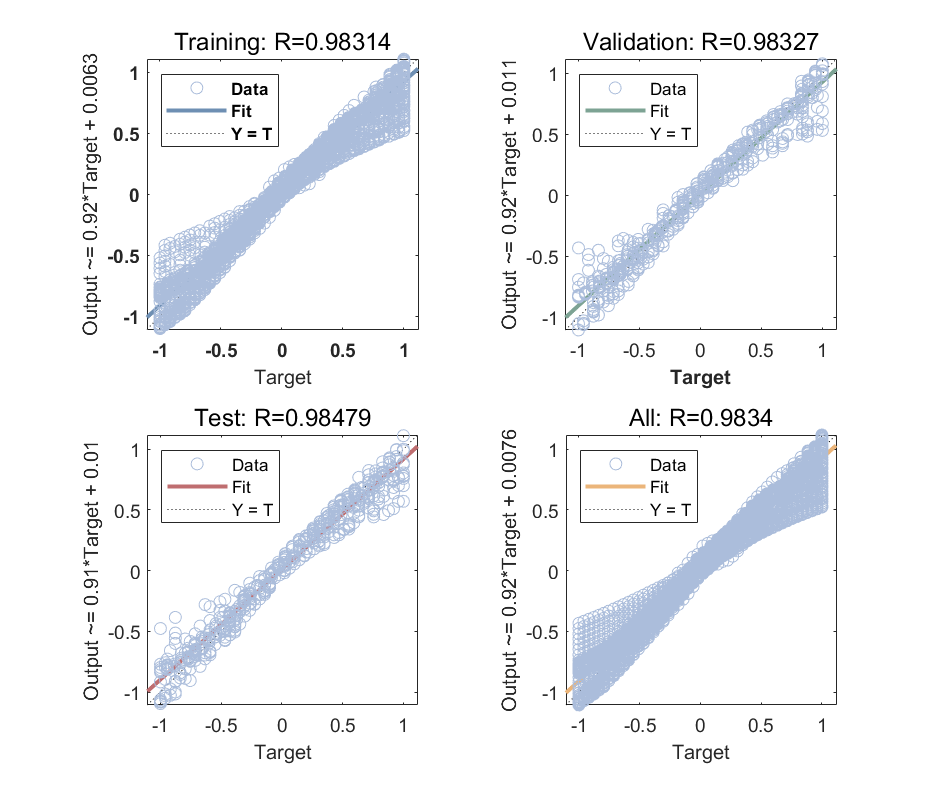}\\
        \hypertarget{fig:8c}{(c)}
\end{minipage}
    \caption{Performance evaluation of the BPNN model over 8 epochs. (a) MSE plot for the training, validation, and test sets, with best validation performance at epoch 7. (b) Gradient magnitude during training, showing changes across epochs. (c) Regression analysis for training, validation, and test sets, presenting the relationship between predicted and actual values, with \(R\)-values close to 1.}

    \label{fig:enter-label}
\end{figure}

The performance on the test set demonstrated high accuracy. The CNN model achieved an RMSE of 0.0058205, a mean absolute error (MAE) of 0.0036943, and a mean absolute percentage error (MAPE) of 0.43395\%, indicating that the model accurately predicted the parameters of the photonic crystal waveguide with minimal error. A comparison of the actual and predicted outputs showed a high degree of alignment, especially in the prediction of the air hole radius, where the two values almost completely overlapped, as illustrated in Fig.9\hyperlink{fig:9b}{(b)}. The model also exhibited minimal deviation in the prediction of waveguide width.

The BPNN model similarly performed well on the test set. Across the 7 test samples, the predicted values closely matched the actual values, as depicted in Fig.9\hyperlink{fig:9a}{(a)}. The $R^2$ value for the air hole radius was 0.9978, and for the waveguide width, it was 0.98574. While the BPNN’s performance was slightly lower than that of the CNN, it showed robust generalization ability with small sample data.
\begin{figure}[H]
    \centering
\begin{minipage}{.49\linewidth}
    \centering
    \includegraphics[width=\linewidth]{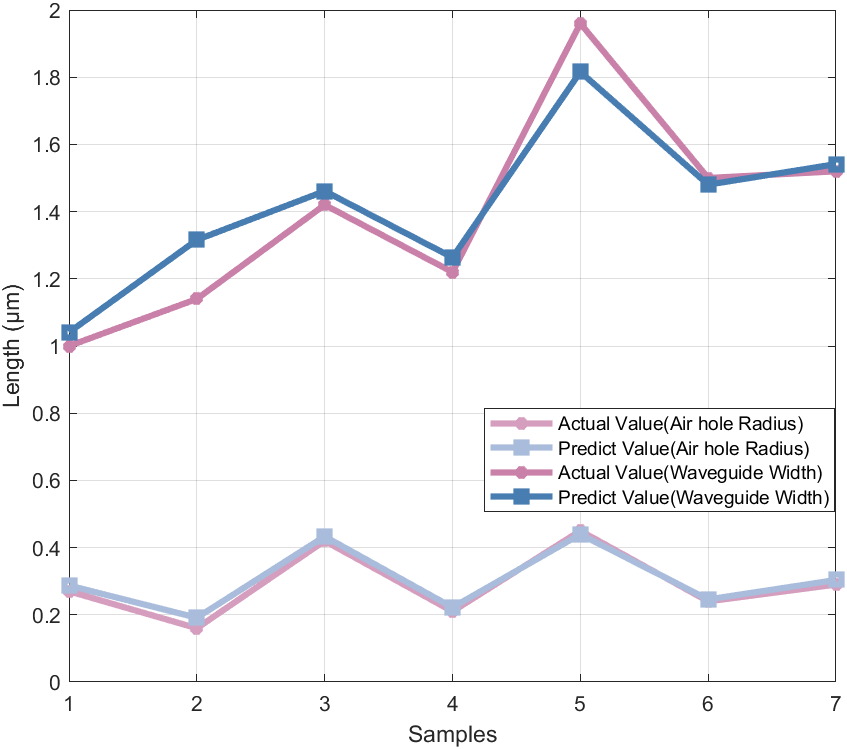}\\
        \hypertarget{fig:9a}{(a)}
\end{minipage}
\begin{minipage}{.49\linewidth}
    \centering
    \includegraphics[width=\linewidth]{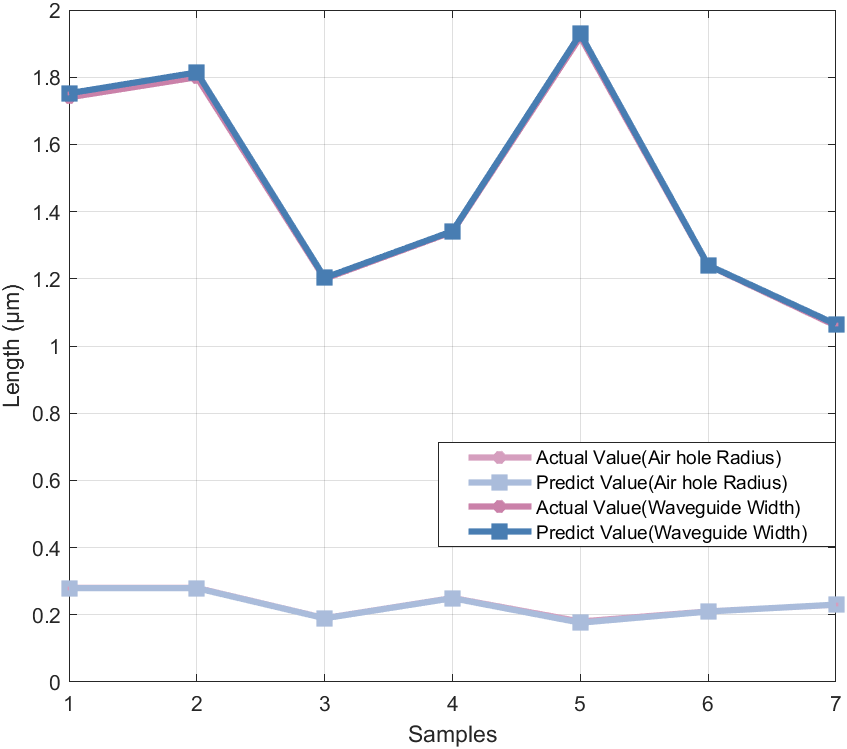}\\
     \hypertarget{fig:9b}{(b)}
\end{minipage}
    \caption{Comparison of predicted and actual values for air hole radius and waveguide width using CNN and BPNN models on the test set. In both plots, the red lines indicate actual values, and the blue lines represent predicted values. The darker lines correspond to waveguide width, and the lighter lines represent air hole radius. (a) Performance of the BPNN model. (b) Performance of the CNN model.}

    \label{fig:enter-label}
\end{figure}

\subsection{CNN and BPNN Comparison}
In terms of training performance, CNN and BPNN exhibited distinct differences. Due to the presence of convolutional layers and a greater number of parameters, the CNN model required longer training times. In contrast, the simpler architecture of BPNN resulted in higher training efficiency, with training typically completed within 8 epochs. Regarding convergence speed, the CNN model demonstrated rapid convergence during the initial epochs but stabilized in later stages. The BPNN model, on the other hand, reached optimal validation loss by the 7th epoch, indicating faster convergence.

As for test set errors, the CNN model produced predictions that were closer to the actual values, particularly in the prediction of waveguide width. Although the BPNN model exhibited slightly larger errors, it still performed well in handling small datasets, displaying strong predictive capability and stability. Both models demonstrated strong generalization abilities, with no indications of overfitting.

From an architectural perspective, both CNN and BPNN have their own advantages. The CNN model, with its convolutional layers, is adept at extracting local geometric features, making it well-suited for handling the spatially dependent structures of photonic crystals. The convolutional layers effectively capture the complex relationships between air hole radius, waveguide width, and dispersion characteristics, while the fully connected layers further process these features to produce accurate predictions. Thus, CNN is particularly effective for tasks involving higher geometric complexity. In contrast, the BPNN model, with its simpler structure, is less capable of capturing spatial features but offers higher training efficiency, shorter training times, and faster convergence. This makes BPNN more suitable for tasks requiring global parameter mapping, especially when working with smaller datasets.

Given the geometric complexity of photonic crystal waveguides, the CNN model’s ability to extract local features through convolutional layers makes it the more favorable choice. With its strong generalization ability and accurate predictive performance, subsequent cross-validation will primarily utilize the CNN model to further ensure its adaptability and precision in handling complex geometric structures.

\subsection{Inverse Design Accuracy with MPB Cross-Validation}
To evaluate the accuracy of the CNN and BPNN models developed in Section 2.4 for inverse design, this study conducted an in-depth analysis of geometric parameter predictions, such as air hole radius and waveguide width. The model predictions were cross-validated through MPB simulations to assess their accuracy. The results of several specific inverse design cases are presented below, along with a discussion on the consistency between the model-predicted geometric parameters and the actual design targets.

In the first experiment, a triangular lattice photonic crystal with a radius of 0.3$\mu \mathrm{m}$ and a waveguide width of $1.0\mu \mathrm{m}$was selected as the test subject. This structure exhibited two guided bands within the bandgap, as shown in Fig.10\hyperlink{fig:10a}{(a)}. Based on this, several modified designs were created to validate the model’s predictive performance. In Experiment 1, the position of the lower guided band was randomly shifted downward, generating a newly artificial designed photonic crystal structure, as illustrated in Fig.10\hyperlink{fig:10b}{(b)}. The structure was then input into the trained CNN model for inverse design. The model output geometric parameters of an air hole radius of $0.3023\mu \mathrm{m}$ and a waveguide width of $1.1551\mu \mathrm{m}$. The model achieved a precision beyond that of the training dataset, with predictions accurate to four decimal places.

\begin{figure}[H]
    \centering
\begin{minipage}{.49\linewidth}
    \centering
    \includegraphics[width=0.9\linewidth]{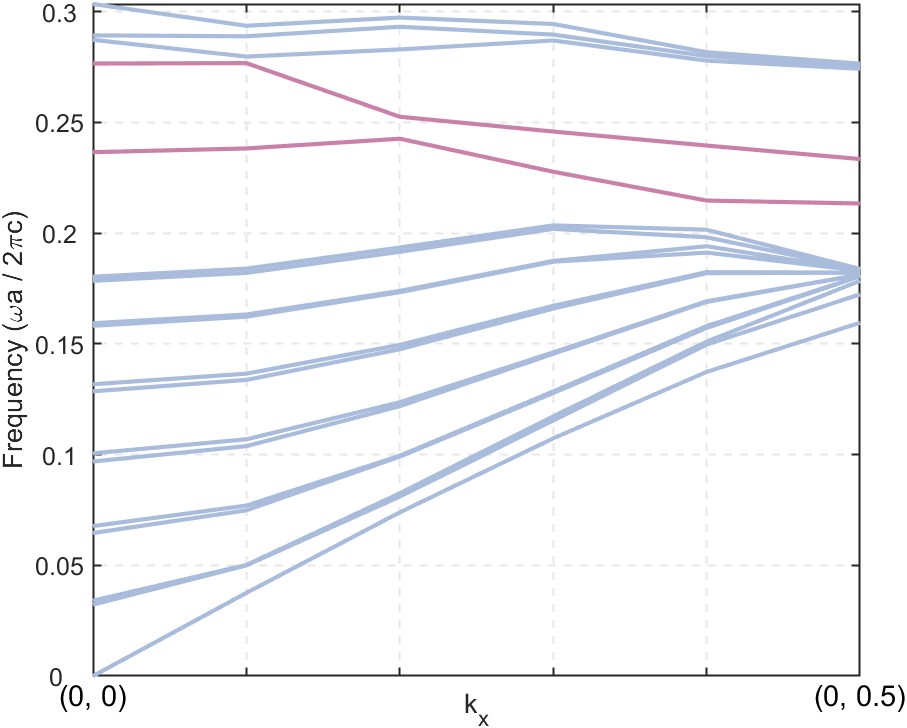}\\
    \hypertarget{fig:10a}{(a)}
\end{minipage}
\begin{minipage}{.49\linewidth}
    \centering
    \includegraphics[width=0.9\linewidth]{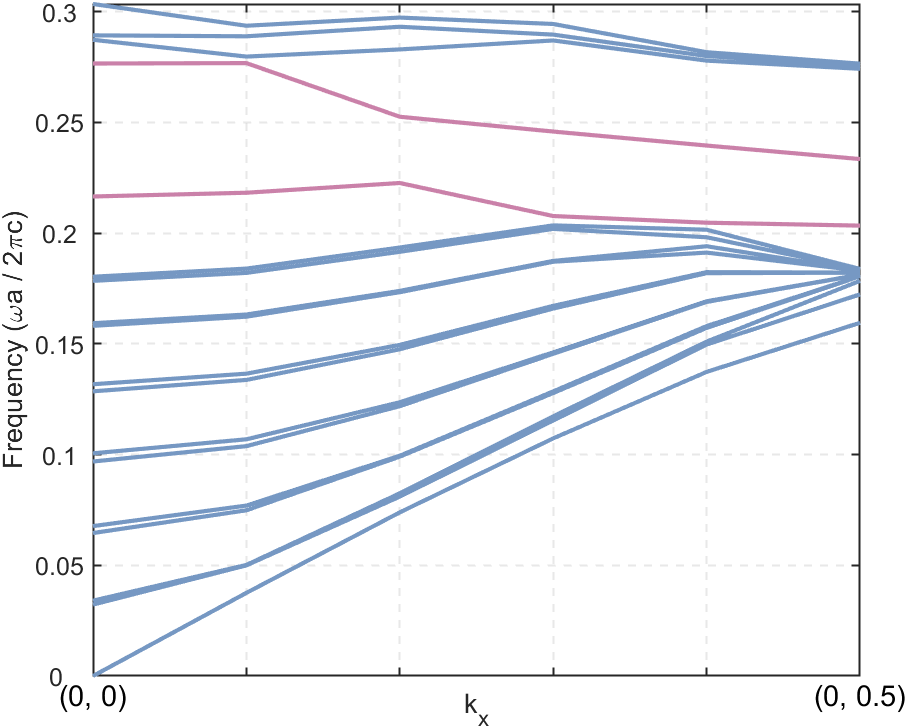}\\
    \hypertarget{fig:10b}{(b)}
\end{minipage}

\begin{minipage}{.49\linewidth}
    \centering
    \includegraphics[width=0.9\linewidth]{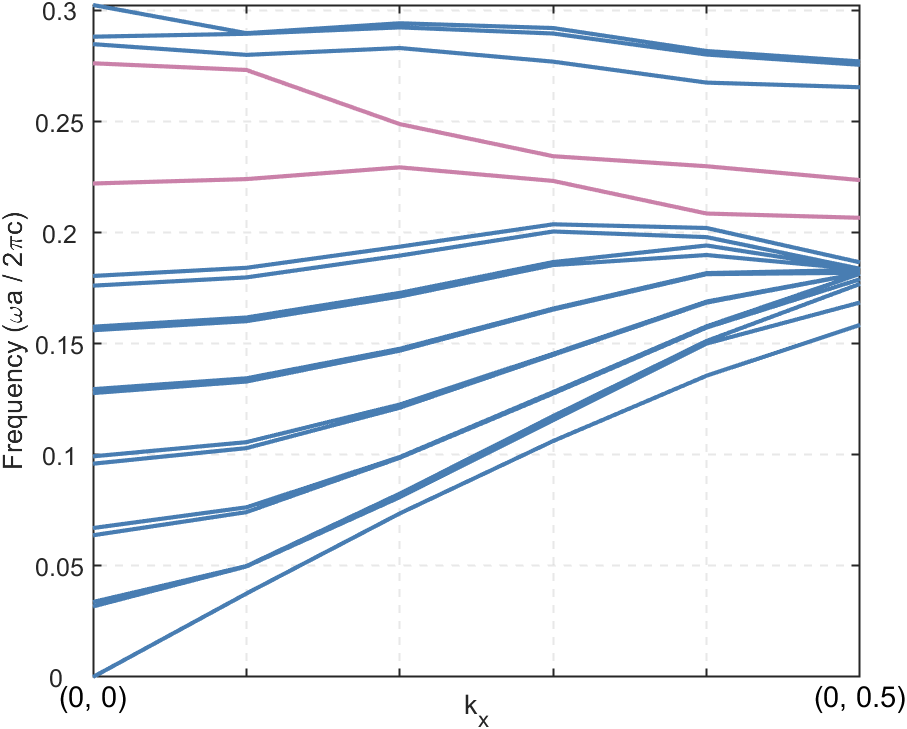}\\
    \hypertarget{fig:10c}{(c)}
\end{minipage}
    \caption{Band diagrams of photonic crystal waveguide designs used to validate the accuracy of the inverse design process in the first experiment. (a) The test subject's photonic band structure with a waveguide width of $1.0\mu \mathrm{m}$ and an air hole radius of $0.3\mu \mathrm{m}$. (b) The artificially designed photonic band structure, with the lower guided band shifted downward. (c) The model-predicted photonic band structure.}
    \label{fig:enter-label}
\end{figure}

The predicted geometric parameters were input into MPB for remodelling, generating a new photonic crystal band structure, as presented in Fig.10\hyperlink{fig:10c}{(c)}. By comparing the artificially designed photonic structure with the new model-generated structure, it was found that the predicted guided band position closely matched the original downward-shifted guided band, and the overall shape of the band structure was very similar to the initial design. This indicates that the CNN model is capable of accurately predicting geometric parameters with high precision.

In the second experiment, the opposite operation was performed by shifting the guided band upward, resulting in a newly artificial designed photonic crystal structure, as shown in Fig.11\hyperlink{fig:11b}{(b)}. This structure was also input into the CNN model for inverse design, and the model predicted geometric parameters of a waveguide width of $0.9384\mu \mathrm{m}$ and an air hole radius of $0.2972\mu \mathrm{m}$. Notably, these parameters exceeded the range set in the original dataset (with the original radius range being $0.15\mu \mathrm{m}$ to $0.49\mu \mathrm{m}$ and the waveguide width range being $1.0\mu \mathrm{m}$ to $1.98\mu \mathrm{m}$). After inputting these parameters into MPB for simulation, the results showed that the lower guided band successfully moved upward,as presented in Fig.11\hyperlink{fig:11c}{(c)},aligning with the intended design target. This further confirmed the effectiveness of the CNN model in inverse design, particularly in predicting new geometric parameters with high precision, even when operating outside the bounds of the original dataset.
\begin{figure}[H]
    \centering
\begin{minipage}{.49\linewidth}
    \centering
    \includegraphics[width=0.9\linewidth]{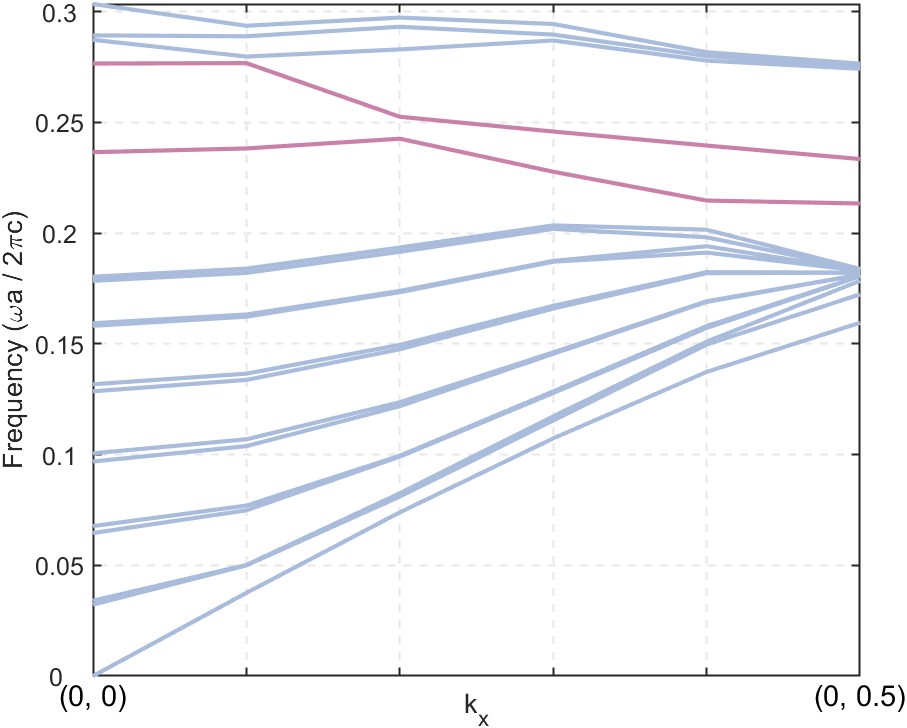}\\
    \hypertarget{fig:11a}{(a)}
\end{minipage}
\begin{minipage}{.49\linewidth}
    \centering
    \includegraphics[width=0.9\linewidth]{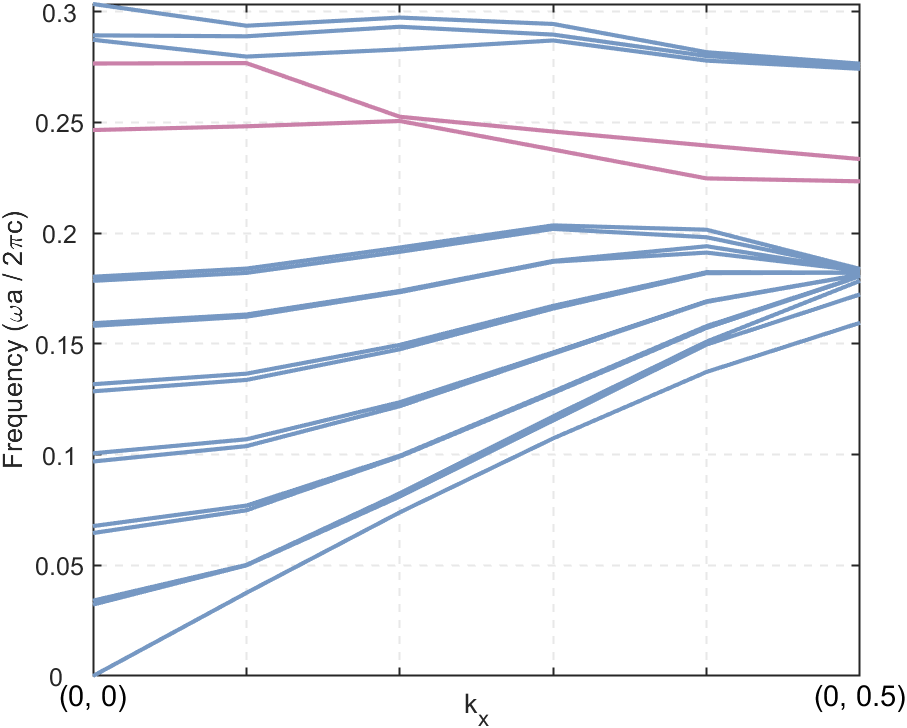}\\
    \hypertarget{fig:11b}{(b)}
\end{minipage}

\begin{minipage}{.49\linewidth}
    \centering
    \includegraphics[width=0.9\linewidth]{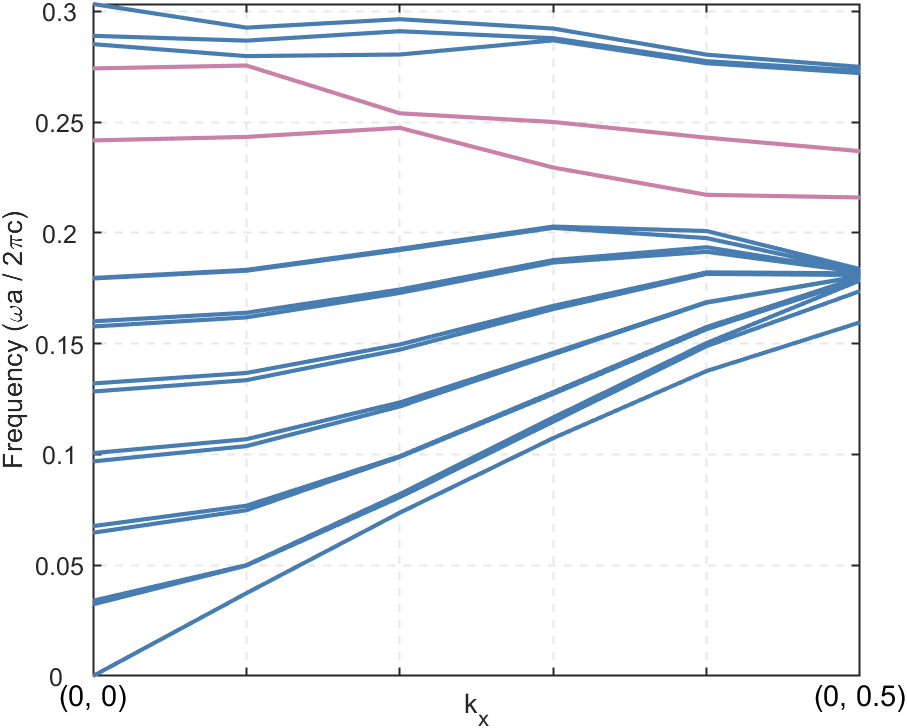}\\
    \hypertarget{fig:11c}{(c)}
\end{minipage}
    \caption{Band diagrams of photonic crystal waveguide designs used to validate the accuracy of the inverse design process in the second experiment. (a) The test subject's photonic band structure with a waveguide width of $1.0\mu \mathrm{m}$ and an air hole radius of $0.3\mu \mathrm{m}$. (b) The artificially designed photonic band structure, with the lower guided band shifted upward. (c) The model-predicted photonic band structure.}
    \label{fig:enter-label}
\end{figure}

In summary, the CNN model performed exceptionally well in inverse design tasks, particularly in maintaining a high degree of predictive accuracy even when working beyond the bounds of the training dataset. The model demonstrated a precision of four decimal places in predicting geometric parameters, and the generated photonic band structure closely matched the original design. The results of the second experiment further demonstrated the robustness and generalization ability of the model, as it was able to effectively predict the corresponding geometric parameters for entirely new design conditions.

Cross-validation through MPB simulations further confirmed the reliability and applicability of the CNN model in the inverse design of photonic crystal waveguides. The model not only demonstrated accurate predictions within the training dataset but also extended its applicability to more complex geometric design tasks, offering a powerful tool for future optimization of photonic crystal structures.

\section{Conclusion}
In this paper, a neural network-based design method was developed to optimize the structural parameters of PCWs.The CNN model demonstrated exceptional predictive accuracy, achieving precision up to four decimal places and delivering stable, reliable predictions even when applied beyond the range of the training data. It significantly outperformed traditional numerical optimization techniques in both accuracy and efficiency. While the BPNN model showed greater computational efficiency on smaller datasets, it encountered limitations when applied to larger and more complex data. Cross-validation with MPB simulations confirmed the CNN model’s ability to accurately predict both lower and upper band edge shifts, aligning closely with the target artificially designed photonic band, further underscoring its versatility across a range of dispersion relations. These results validate the reliability and generalization capability of the CNN model, highlighting its practical value in photonic device design.

The findings demonstrate that deep learning techniques offer a promising approach for the rapid and precise design of PCWs, surpassing traditional methods in both predictive accuracy and computational efficiency. This method provides a powerful tool for the development of low-loss, compact photonic devices, laying a strong foundation for future advancements in integrated photonics and optical communication systems.


\bibliography{sample}

\begin{thebibliography}{10}
\newcommand{\enquote}[1]{``#1''}

\bibitem{Joannopoulos:08:Book}
J.~D. Joannopoulos, S.~G. Johnson, J.~N. Winn, and R.~D. Meade, \emph{Photonic Crystals: Molding the Flow of Light (Second Edition)} (Princeton University Press, 2008), 2nd ed.

\bibitem{2}
E.~{Yoxall}, M.~{Schnell}, A.~Y. {Nikitin}, \emph{et~al.}, \enquote{{Direct observation of ultraslow hyperbolic polariton propagation with negative phase velocity},} {\protect\JournalTitle{Nature Photonics}} \textbf{9}, 674--678 (2015).

\bibitem{3}
M.~K. Moghaddam and R.~Fleury, \enquote{Slow light engineering in resonant photonic crystal line-defect waveguides,} {\protect\JournalTitle{Opt. Express}} \textbf{27}, 26229--26238 (2019).

\bibitem{4}
H.~GOTO, Y.~TSUJI, T.~YASUI, and K.~HIRAYAMA, \enquote{A study on optimization of waveguide dispersion property using function expansion based topology optimization method,} {\protect\JournalTitle{IEICE Transactions on Electronics}} \textbf{E97.C}, 670--676 (2014).

\bibitem{5}
M.~A. Foster, K.~D. Moll, and A.~L. Gaeta, \enquote{Optimal waveguide dimensions for nonlinear interactions,} {\protect\JournalTitle{Opt. Express}} \textbf{12}, 2880--2887 (2004).

\bibitem{6}
L.~Zhang, Y.~Yue, Y.~Xiao-Li, \emph{et~al.}, \enquote{Flat and low dispersion in highly nonlinear slot waveguides,} {\protect\JournalTitle{Opt. Express}} \textbf{18}, 13187--13193 (2010).

\bibitem{7}
S.~Xia, D.~Jukić, N.~Wang, \emph{et~al.}, \enquote{Nontrivial coupling of light into a defect: the interplay of nonlinearity and topology,} {\protect\JournalTitle{Light: Science \& Applications}} \textbf{9}, 147 (2020).

\bibitem{8}
Y.~Bidaux, F.~Kapsalidis, P.~Jouy, \emph{et~al.}, \enquote{Coupled-waveguides for dispersion compensation in semiconductor lasers,} {\protect\JournalTitle{Laser \& Photonics Reviews}} \textbf{12}, 1700323 (2018).

\bibitem{9}
M.~Notomi, K.~Yamada, A.~Shinya, \emph{et~al.}, \enquote{Extremely large group-velocity dispersion of line-defect waveguides in photonic crystal slabs,} {\protect\JournalTitle{Phys. Rev. Lett.}} \textbf{87}, 253902 (2001).

\bibitem{10}
V.~D.~R. {Pavan}, V.~{Nikhil}, K.~{Dey}, \emph{et~al.}, \enquote{{Analysing group indices and dispersion characteristics of engineered photonic crystal waveguides using artificial neural network},} {\protect\JournalTitle{Journal of Optics}} \textbf{53}, 1438--1446 (2023).

\bibitem{11}
A.~Pradhan, C.~Prakash, T.~Datta, \emph{et~al.}, \enquote{Ann-based estimation of dispersion characteristics of slotted photonic crystal waveguides,} {\protect\JournalTitle{Journal of Computational Electronics}} \textbf{23}, 1--9 (2024).

\bibitem{12}
S.~Molesky, Z.~Lin, A.~Piggott, \emph{et~al.}, \enquote{Inverse design in nanophotonics,} {\protect\JournalTitle{Nature Photonics}} \textbf{12}, 659--670 (2018). Publisher Copyright: {\textcopyright} Springer Nature Limited 2018.

\bibitem{13}
P.~R. Wiecha, A.~Arbouet, C.~Girard, and O.~L. Muskens, \enquote{Deep learning in nano-photonics: inverse design and beyond,} {\protect\JournalTitle{Photon. Res.}} \textbf{9}, B182--B200 (2021).

\bibitem{14}
M.~Minkov, I.~A.~D. Williamson, L.~C. Andreani, \emph{et~al.}, \enquote{Inverse design of photonic crystals through automatic differentiation,} {\protect\JournalTitle{ACS Photonics}} \textbf{7}, 1729–1741 (2020).

\bibitem{15}
K.~Y. Yang, J.~Skarda, M.~Cotrufo, \emph{et~al.}, \enquote{Inverse-designed non-reciprocal pulse router for chip-based lidar,} {\protect\JournalTitle{Nature Photonics}} \textbf{14}, 369–374 (2020).

\bibitem{16}
A.~Y. Piggott, J.~Lu, K.~G. Lagoudakis, \emph{et~al.}, \enquote{Inverse design and demonstration of a compact and broadband on-chip wavelength demultiplexer,} {\protect\JournalTitle{Nature Photonics}} \textbf{9}, 374–377 (2015).

\bibitem{17}
Q.~Zhuge, X.~Zeng, H.~Lun, \emph{et~al.}, \enquote{Application of machine learning in fiber nonlinearity modeling and monitoring for elastic optical networks,} {\protect\JournalTitle{J. Lightwave Technol.}} \textbf{37}, 3055--3063 (2019).

\bibitem{18}
X.~Tu, W.~Xie, Z.~Chen, \emph{et~al.}, \enquote{Analysis of deep neural network models for inverse design of silicon photonic grating coupler,} {\protect\JournalTitle{Journal of Lightwave Technology}} \textbf{PP}, 1--1 (2021).

\bibitem{19}
Y.~Song, D.~Wang, J.~Qin, \emph{et~al.}, \enquote{Physical information-embedded deep learning for forward prediction and inverse design of nanophotonic devices,} {\protect\JournalTitle{Journal of Lightwave Technology}} \textbf{PP}, 1--1 (2021).

\bibitem{20}
R.~E. Christiansen and O.~Sigmund, \enquote{Inverse design in photonics by topology optimization: tutorial,} {\protect\JournalTitle{JOURNAL OF THE OPTICAL SOCIETY OF AMERICA B-OPTICAL PHYSICS}} \textbf{38}, 496--509 (2021).

\bibitem{21}
Y.~Chen, J.~Qiu, D.~Zhenli, \emph{et~al.}, \enquote{Inverse design of free-form devices with fabrication-friendly topologies based on structure transformation,} {\protect\JournalTitle{Journal of Lightwave Technology}} \textbf{PP}, 1--15 (2023).

\bibitem{22}
L.~Su, A.~Y. Piggott, N.~V. Sapra, \emph{et~al.}, \enquote{Inverse design and demonstration of a compact on-chip narrowband three-channel wavelength demultiplexer,} {\protect\JournalTitle{ACS PHOTONICS}} \textbf{5}, 301--305 (2018).

\bibitem{23}
S.~Molesky, Z.~Lin, A.~Piggott, \emph{et~al.}, \enquote{Inverse design in nanophotonics,} {\protect\JournalTitle{Nature Photonics}} \textbf{12}, 659--670 (2018). Publisher Copyright: {\textcopyright} Springer Nature Limited 2018.

\bibitem{24}
R.~E. Christiansen and O.~Sigmund, \enquote{Compact 200 line matlab code for inverse design in photonics by topology optimization: tutorial,} {\protect\JournalTitle{JOURNAL OF THE OPTICAL SOCIETY OF AMERICA B-OPTICAL PHYSICS}} \textbf{38}, 510--520 (2021).

\bibitem{25}
Y.~Chen, Y.~Hu, J.~Zhao, \emph{et~al.}, \enquote{Topology optimization-based inverse design of plasmonic nanodimer with maximum near-field enhancement,} {\protect\JournalTitle{ADVANCED FUNCTIONAL MATERIALS}} \textbf{30} (2020).

\end{thebibliography}






\end{document}